\documentclass{PoS}

\title{Staggered Dslash Performance on Intel Xeon Phi Architecture}

\ShortTitle{Staggered Dslash on Xeon Phi}

\author{\speaker{Ruizi Li}\thanks{Current address: Department of Physics,
University of Wuppertal, D-42119 Wuppertal, Germany.}\\
        Department of Physics, Indiana University, Bloomington IN 47405, USA\\
        E-mail: \email{ruizli AT umail.iu.edu}}

\author{Steven Gottlieb\\
        Department of Physics, Indiana University, Bloomington IN 47405, USA\\
        E-mail: \email{sg AT iub.edu}}

\abstract{The conjugate gradient (CG) algorithm is among the most essential 
and time consuming parts of lattice calculations with staggered quarks. 
We test the performance of CG and dslash, the key step in the CG algorithm, 
on the Intel Xeon Phi, also known as the Many Integrated Core (MIC) 
architecture.  We try different parallelization strategies using MPI,
OpenMP, and the vector processing units (VPUs).}

\FullConference{The 32nd International Symposium on Lattice Field Theory,\\
		23-28 June, 2014\\
		Columbia University New York, NY}

\begin{document}

\section{Introduction}
As we aspire to compute at the exascale level, it is clear that power
consumption is going to be an essential issue \cite{kogge08}.  
For the past several years,
GPU computing has been explored and exploited to achieve high levels of
performance and energy efficiency 
\cite{Barros:2008rd,Babich:supercomputing,Clark:2009wm,Shi2010a,Gottlieb:2010zz,Babich:2011zz,Babich:2011np}.  
Intel has been developing and
selling many-core chips that have recently been appearing in computers in the 
Top 500 list \cite{top500}.  The Many Integrated Core (MIC) architecture uses
simple, energy efficient cores.
Intel Xeon Phi processors, based on this architecture, started shipping
in late 2012.  We have been part of the Beacon Project at 
the National Institute of Computational Sciences at The University
of Tennessee.  Beacon, which contains both Intel Xeon and Xeon Phi processors
\cite{beaconweb},
achieved the top spot on the November, 2012 Green500 \cite{green500}.

In the next section, we provide some background on the Xeon Phi architecture
and the options for using it.  In Sec.~3, we describe our experience
using MPI and a combination of MPI and OpenMP with the MILC code.
Section 4 describes our experience using vectorization.  We have done that
both with the normal MILC layout and with a data layout 
devised originally for Wilson fermions \cite{JooISC}.
Finally, we offer some conclusions.

\section{Xeon Phi Background}
The nodes of Beacon \cite{beaconweb} contain two 8-core Intel 
Xeon E5-2570 and four Xeon Phi 5110P coprocessors.  
Each coprocessor contains 8 GB of high bandwidth GDDR5 memory.
The host node also contains 256 GB of DDR3 memory.  
The current generation of Xeon Phi coprocessors are  
named Knights Corner.  
The 5110P coprocessor has 60 cores and runs at a
speed of 1.053 GHz.  There are other Xeon Phi coprocessors with 61 cores
running at 1.238 GHz and containing 16 GB of memory, as well as coprocessors
with only 57 cores and 6 GB of memory.

Each Xeon Phi core has a 512 KB L2 cache and separate 32 KB data and 
instruction L1 caches.  Each core is capable of supporting four hardware
threads and cannot reach maximum performance without at least two threads
assigned to each core.  The bulk of the floating point power of the
Xeon Phi processor comes from the 512 bit wide SIMD floating point unit
that is capable of performing 8 double precision or 16 single precision
SIMD operations simultaneously.  The Xeon Phi cores and memory controllers
are connected in a ring.  The guaranteed not to exceed memory bandwidth of the
coprocessor GDDR5 memory is 353 GB/s.  Achieving something close to half 
that rate may be possible for well written code.  Access to host memory
goes through the PCI bus and is much slower.
In double precision, the SIMD vector unit which can do a fused 
multiply-add (FMA) has a peak speed of about 1 TF/s.  Using half the
peak bandwidth, we expect about 0.174 bytes/flop are available for
double precision operations.  If one does not try to use the vector unit,
the ratio would be 1.4 bytes/flop.  Since ordinary arithmetic operations
such as add and multiply require 16 bytes of input and 8 bytes of 
output per flop, it is clear that one needs to reuse data that can remain
in cache.

Current computers with Xeon Phi coprocessors allow a number of different
styles of programming.  In offload mode, the code begins running on the
Xeon processors and parallel sections of code are directed to the
coprocessors using compiler directives (\textit{e.g.}, \texttt{pragma} 
statements).  
Data may need to be explicitly
copied to the Xeon Phi memory.  In native mode, the code is cross
compiled for the MIC architecture, and is launched to run directly 
on the coprocessors.  This is the
only mode we have explored.  It is also possible to try to use both
Xeon and Xeon Phi cores by appropriately breaking up the calculation, or
by running two separate jobs.

The nodes of the Beacon computer are connected with FDR Infiniband capable of
a peak of 56 Gb/s bi-directional bandwidth on each link.

\section{MILC with MPI and OpenMP}
One aspect of Xeon Phi programming is that it should be very easy to
recompile code that runs on Xeon processors to run on the Xeon Phi.  However,
as we shall see, the performance may not be that impressive.  The MILC code
was originally written to run with a variety of message passing APIs, but
for many years MPI \cite{MPI}
and QMP \cite{QMP} have been the most important options.  In 2000,
we experimented with a hybrid combination of OpenMP and MPI 
\cite{Gottlieb:2000tn}.  However,
at that time we did not notice a significant improvement from the 
combination and dropped the OpenMP pragmas.  Motivated by the 
Xeon Phi, we have again added OpenMP pragmas to the code.

Starting from MILC version 7.7.8, we have benchmarked the CG solver for
HISQ configuration generation in native mode on Beacon.  We considered a
multishift solver, as it is an important part of the gauge generation code
using the Rational Hybrid Molecular Dynamics or Rational Hybrid Monte Carlo 
algorithms.  We used either 9 or 11 shifts, but only about 50 iterations
were required for convergence in our benchmark runs.  Using a pure MPI
code the performance was between 12 and 16 GF/s on a single coprocessor.
Using a combination of OpenMP and MPI, performance increases to 15 to 20
double precision GF/s.

The multishift solver uses a large number of SU(3) vectors, both for the
solutions and temporaries.  
Each double precision complex 3-vector takes up $3\times 2\times8=48$ bytes.
Each $3\times3$ matrix takes up three times that amount.  Letting
$N_m$ be the number of masses, and $V$ be the number of grid points,
the memory usage is $(36 + 2 N_m) 48 V$ bytes,
which includes the 1-link and 3-link matrices and all the SU(3) vectors
used in the computation, but ignores the index vectors used for gathers.
In our benchmarks $N_m=9$ or 11.  This amounts to 2592 or 2784 bytes per
grid point.
The number of operations per CG iteration is $(1205+15 N_m) V$.
The bulk of the floating point operations are in dslash, for which
we will calculate the arithmetic intensity.
For each grid point, the improved staggered dslash must access 16 $3\times3$ 
complex matrices.  For each of the 8 (positive and negative) directions both
1-link and 3-link matrices are needed.  We also need 16 3-component complex
vectors.  The output is a single 3-component complex vector.  The operations
are 16 matrix-vector multiplies and 15 vector adds.  Each matrix-vector
multiply requires 36 real multiplies and 30 real adds.  Each vector add
requires 6 real adds.  Thus, at each grid point, there are 576 real
multiplies and 576 real adds or 1146 flops.  There are 384 operands of 
input data.  This amounts to 2.98 flops per operand.  That is 0.75 flops/byte
in single precision or 0.37 flops per byte in double precision.  The MILC
code makes use of a number of temporary vectors not counted above, which would
reduce the arithmetic intensity.
The ratio peak flops to bandwidth of the Xeon Phi chip is close to three, so
it is clear that efficient use of bandwidth and cache reusage will be important
for code optimization.

Carleton DeTar prepared a test code for benchmarking the single mass CG
inverter with OpenMP.  With 600 iterations to convergence, he was getting
about 42 GF/s on a Xeon Phi
coprocessor of Stampede at the Texas Advanced Computing Center.  Performance
slightly increased to about 45 GF/s when running over 1000 iterations.
Stampede contains (pre-production) Xeon Phi SE10P coprocessors that have
a peak speed of 1.07 TF/s.

It is clear that without an effort to vectorize the code, the MILC
performance is only 2--4.5 \% of the peak speed of the Xeon Phi.

\section{Vectorization}
Our vectorization efforts for staggered quarks are based on the 
pioneering work with Wilson quarks \cite{JooISC}.
The longest vector length in QCD codes is obviously the number of grid points.
However, as we are dealing with a SIMD vector unit, not the type of 
vector unit on early Cray computers that could do gather/scatter at full
speed, we need to be very careful about the data layout.  In single precision,
the SIMD vector length $VECLEN$ is 16.  Because of the even-odd preconditioning 
in the solver, one possibility is to have the lattice dimension (at least in
one direction) be a multiple of 32.  However, this is rather restrictive.
Thus, it was decided to write a code in which variables from
multiple sites in the $x$-direction are stored contiguously.  This
strategy is often called {\it structure of arrays}.  The code supports
$SOALEN=4$, 8, and 16.  However, since $SOALEN$ can be $<16$, we need to
aggregate values with different $y$ coordinates to fill the vector
registers.  Thus acceptable grid sizes must be a multiple of $SOALEN$
in the $x$ direction and a multiple of $VECLEN/SOALEN$ in the $y$-direction.
Further details may be found in Ref.~\cite{JooISC}.

\begin{figure}[ht]
\centering
\includegraphics[scale=0.45]{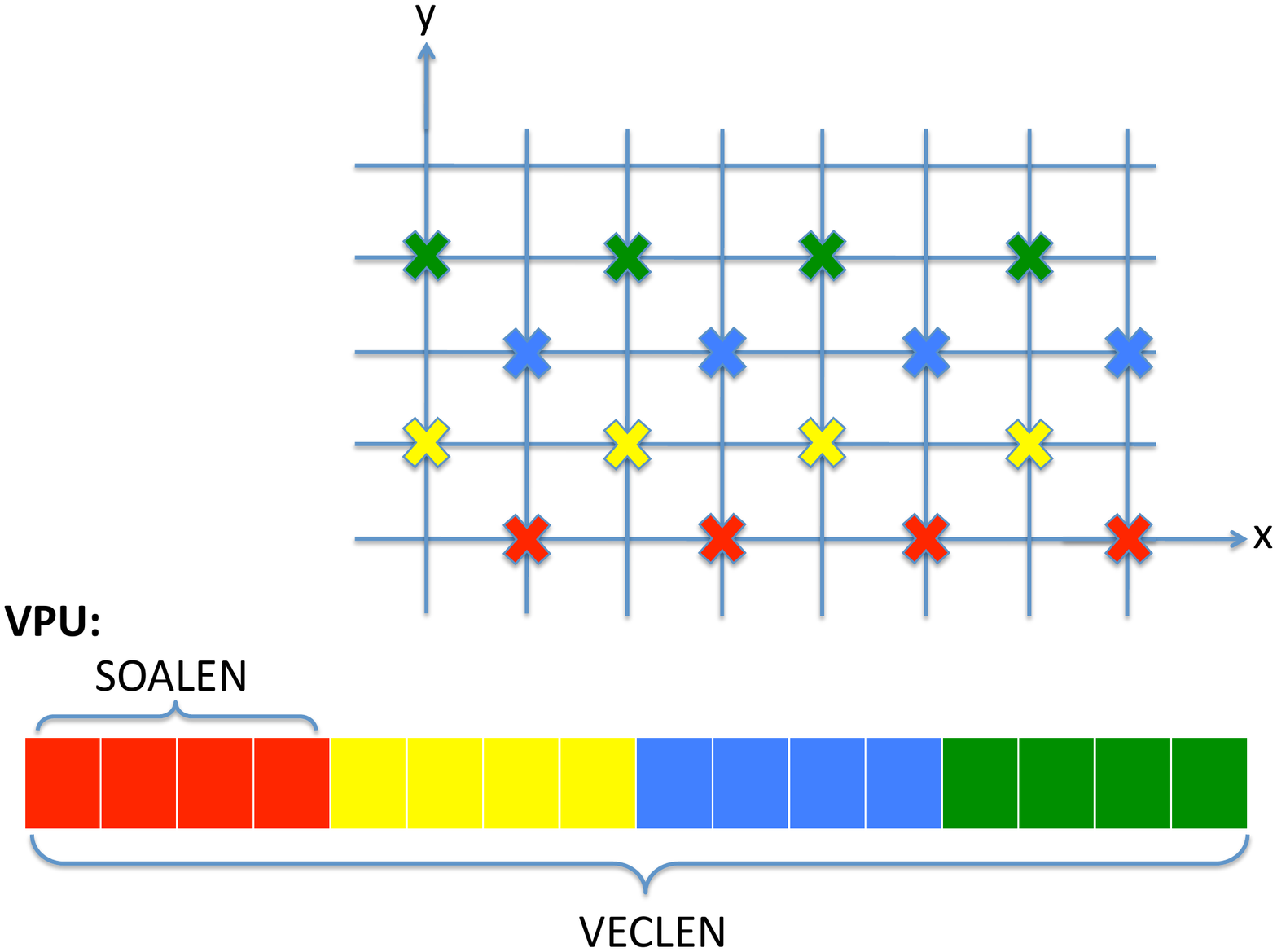}
\caption{With $SOALEN=4$ data from four consecutive sites in the $x$-direction
are combined another three groups of four sites with the same parity and
larger $y$ coordinates to fill a vector register in the vector processing 
unit (VPU).}
\label{fig:layout}
\end{figure}

Figure~\ref{fig:layout}
shows an $x$-$y$ plane of the lattice and how points from the
grid are aggregated into vector registers.  In this case, $SOALEN=4$.
The MILC data layout for an SU(3) vector is {\tt complex KS[3]}.
In the current code, the data is stored in memory 
as {\tt float KS[3][2][$SOALEN$]}, where 3 is for color and
2 is for real and imaginary parts.  In registers, the data from
different $y$ values is aggregated into the form {\tt float KS[3][2][$VECLEN$]}.
This layout is very similar to that used for Wilson quarks where there
is an additional spinor index appearing before the color index.

The gauge fields in the MILC code are arrays of the form {\tt complex
Gauge[4][3][3]}.  In the Xeon Phi code, we allow for compression of the
gauge field and we pack sites together in groups of $SOALEN$ or $VECLEN$.
Thus the array definitions are
{\tt Gauge[8][$GROWS$][3][2][$SOALEN$]} or 
{\tt Gauge[8][$GROWS$][3][2][$VECLEN$]} where $GROWS$ is 3 for uncompressed
storage and 2 if one wants to reconstruct the third row from the first two.
In this code, both forward and backward links are stored at each site to 
increase locality in memory.  This avoids referencing backward links from
four neighboring sites.

The performance study is based on a ``micro benchmark'' test code that is
not part of a full evolution code.  Starting from the Wilson code, it was 
modified to do naive staggered quarks (ignoring the Kogut-Susskind
phase factors that are normally absorbed into the gauge fields).  Later, the
phase factors were added and finally, the three-link term was added.
The code generator for Wilson quarks was modified to generate appropriate
intrinsics for staggered operations and for gathering third-nearest
neighbors needed for the Naik term.  We investigated gauge compression,
although with smearing reconstructing the third row from the
first two will not suffice for the one-link term.  Two lattice grids
were used in these tests:  $32^3\times 128$ and $32\times40\times24\times96$.
A technique to improve speed of storing the result of dslash in memory,
called streaming stores, was investigated.  It avoids a read, modify,
write cycle for storing data.  We used this technique with $SOALEN=8$ and 16.
(With $SOALEN=4$, stores are not always aligned on cache line boundaries,
precluding the use of streaming stores.  This would not be so in double
precision.)
In addition to measuring the rate of floating point operations, we estimate
the sustained bandwidth to coprocessor memory.

Without gauge field compression, our staggered dslash performance is
138--141 GF/s, and estimated bandwidth is 152--159 GB/s.  With gauge
compression, the performance increases to 184--193 GF/s, and the bandwidth
is reduced to 142--153 GB/s.  (The numbers we quote here cover a narrower
range than those in Table 1   
as more variation was seen with the other grid size.)
Including the staggered phases in the 
reconstruction of the compressed gauge fields did not effect performance
very much.  It ranged from 179--196 GF/s and the bandwidth estimate was
virtually unchanged.  We generally found small increases in performance 
for larger values of $SOALEN$.

\begin{table}[b]
\begin{center}
\begin{tabular}{cccccc}\hline\hline
Streaming store&$SOALEN$&Uncompressed&Uncompressed&Compressed&Compressed\\
&&(GF/s)&(GB/s)&(GF/s)&(GB/s)\\
\hline\hline
No&4&139&158&184&147\\
No&8&140&159&188&150\\
No&16&139&158&189&152\\
Yes&8&139&153&187&142\\
Yes&16&140&154&190&144\\
\hline\hline
\end{tabular}
\label{table:naive}
\caption{Performance of naive staggered dslash code for a $32^3\times 128$ grid
on a single Xeon Phi coprocessor.}
\end{center}
\end{table}
\begin{table}[tb]
\begin{center}
\begin{tabular}{cccccc}\hline\hline
Streaming store&$SOALEN$&Uncompressed&Uncompressed&Compressed&Compressed\\
&&(GF/s)&(GB/s)&(GF/s)&(GB/s)\\
\hline\hline
No&4&136&145&179&131\\
No&8&139&149&192&141\\
No&16&135&144&193&142\\
Yes&8&140&146&193&137\\
Yes&16&140&147&193&137\\
\hline\hline
\end{tabular}
\label{table:withnaik}
\caption{Performance of dslash code including Naik term for a $32^3\times 128$ grid
on a single Xeon Phi coprocessor.}
\end{center}
\end{table}

Adding the Naik term to the quark action does not result in a large
change in performance.  Without gauge compression, we see 136--140 GF/s
and an estimated bandwidth of 144--149 GB/s.  Compressing both the
one-link and three-link terms, we see 178-193 GF/s and bandwidth of
130-141 GB/s.  (See Table 2.)  
A real calculation would either use compression only
on the three-link term or implement 9-element compression for the
smeared links as they are no longer elements of SU(3).

Code was required to calculate the three-link products.  Since this
involves SU(3) matrix-matrix, rather than SU(3) matrix-vector operations,
the arithmetic intensity is higher and so is the performance.  Without
gauge compression, performance is about 184 GF/s, requiring a bandwidth of
134 GB/s.  Turning on compression, the performance is about 304 GF/s and
bandwidth estimate is 147 GB/s.  It might be possible to tune prefetching
for the uncompressed case to improve bandwidth and, hence, performance.

We also tried to vectorize the naive dslash code with the MILC data layout.
Performance was about 50 GF/s.

\section{Conclusions}
We have been developing code for staggered quarks on the Xeon Phi coprocessor
and find that with a well designed data layout we can achieve about 185
GF/s in single precision with compression of gauge matrices and about
140 GF/s without compression.  Having an appropriate data layout is
critical to performance because it is needed to exploit the SIMD vector
unit.  Memory bandwidth is a critical limitation on code performance.  For
staggered quarks, we did not find much dependence on $SOALEN$ or whether
or not we use streaming stores.

We have also attempted to use MPI across multiple coprocessors, but that
has not met with much success.  The standard implementation of MPI currently
has limited performance.  
There are other implementations
of MPI that we would like to try.  Although substantial progress has
been made, much work remains to produce a production-ready staggered code
capable of generating configurations.

\medskip
{\bf Acknowledgments:}
This material is based on work supported by the 
National Science Foundation under Grant Number 1137097 and by the 
University of Tennessee through the Beacon Project. 
Any opinions, findings, conclusions, or recommendations expressed 
in this material are those of the authors and do not necessarily reflect 
the views of the National Science Foundation or the University of Tennessee.
We are grateful to Glenn Brook, the National Institute of 
Computational Sciences, and The University of Tennessee for including
us in the Beacon Project.

We are also very grateful for the extremely helpful engagement of
B\'alint Jo\'o, and his Intel collaborators
Dhiraj D.~Kalamkar, Mikhail Smelyanskiy, Karthikeyan Vaidyanathan, 
and others, upon whose work all the vectorization work was based.
We appreciate many useful suggestions on the manuscript from
B\'alint Jo\'o.

This work was supported by DOE grants FG02-91ER 40661 and DE-SC0010120,
and by the NSF/University of Tennessee Award A12-0848-S004.

\end{document}